\documentclass{article}
\usepackage{amsmath,amssymb,amsfonts,amsthm}
\usepackage[russian]{babel}
\usepackage{color}
\usepackage{graphicx}
\usepackage{wrapfig}

\begin{document}

\flushbottom

\renewcommand{\figurename}{Fig.}
\def\refname{References}
\def\proofname{Proof}

\newtheorem{teo}{Theorem}
\newtheorem{pro}{Proposition}
\newtheorem{rem}{Remark}

\def\tens#1{\ensuremath{\mathsf{#1}}}

\if@mathematic
   \def\vec#1{\ensuremath{\mathchoice
                     {\mbox{\boldmath$\displaystyle\mathbf{#1}$}}
                     {\mbox{\boldmath$\textstyle\mathbf{#1}$}}
                     {\mbox{\boldmath$\scriptstyle\mathbf{#1}$}}
                     {\mbox{\boldmath$\scriptscriptstyle\mathbf{#1}$}}}}
\else
   \def\vec#1{\ensuremath{\mathchoice
                     {\mbox{\boldmath$\displaystyle\mathbf{#1}$}}
                     {\mbox{\boldmath$\textstyle\mathbf{#1}$}}
                     {\mbox{\boldmath$\scriptstyle\mathbf{#1}$}}
                     {\mbox{\boldmath$\scriptscriptstyle\mathbf{#1}$}}}}
\fi

\begin{center}
{\Large\bf Figures of equilibrium of an inhomogeneous self-gravitating
fluid\\}

\bigskip

\footnotetext{The work of Alexey V.\,Borisov was carried out within the
framework of the state assignment to the Udmurt State University ``Regular
and Chaotic Dynamics''. The work of Ivan S.\,Mamaev was supported by the
RFBR grant 14-01-00395-a. The work of Ivan A.\,Bizyaev was supported by
the RFBR grant 13-01-12462-ofi\_m.}

{\large\bf Ivan~A.\,Bizyaev$^1$,
Alexey~V.\,Borisov$^2$,
Ivan~S.\,Mamaev$^3$\\}
\end{center}

\begin{quote}
\begin{small}
\noindent
$^1$ Udmurt State University, Universitetskaya 1, Izhevsk, 426034,
Russia\\
$^2$ National Research Nuclear University ``MEPhI'', Kashirskoye shosse 31, Moscow, 115409,
Russia. E-mail: borisov@rcd.ru\\
$^3$ Institute of Mathematics and Mechanics of the Ural Branch of
RAS, S.\,Kovalevskaja str. 16, Ekaterinburg, 620990,  Russia
\end{small}

\bigskip
\bigskip

\begin{small}
\textbf{Abstract.} This paper is concerned with the figures of equilibrium
of a~self-gravitating ideal fluid with density stratification and
a~steady-state velocity field. As in the classical setting, it is assumed
that the figures or their layers uniformly rotate about an axis fixed in
space.

It is shown that the ellipsoid of revolution (spheroid) with
confocal\linebreak stratification, in which each layer rotates with
inherent constant angular velocity, is at equilibrium. Expressions are
obtained for the gravitational potential, change in the angular velocity
and pressure, and the conclusion is drawn that the angular velocity on the
outer surface is the same as that of the Maclaurin spheroid. We note that
the solution found generalizes a~previously known solution for piecewise
constant density distribution. For comparison, we also present a~solution,
due to Chaplygin, for\linebreak a~homothetic density stratification.

We conclude by considering a~homogeneous spheroid in the space of constant
positive curvature. We show that in this case the spheroid cannot rotate
as a~rigid body, since the angular velocity distribution of fluid
particles depends on the distance to the symmetry axis.

\smallskip


\smallskip

\end{small}
\end{quote}

\clearpage

\section*{Introduction}

This paper is concerned with exact solutions to the problem
of (axisymmetric) figures of equilibrium of a~self-gravitating ideal fluid
with density \textit{stratification}. First of all, we briefly recall the
well-known results along these lines.

For a~\textit{homogeneous} fluid, the following ellipsoidal equilibrium
figures are well known for which the entire mass \textit{uniformly rotates
as a~rigid body} about a~fixed axis:
\begin{itemize}
\itemsep=-1pt
\item[] the Maclaurin spheroid (1742),

\item[] the Jacobi ellipsoid (1834),
\end{itemize}

In addition, in the case of a~homogeneous fluid there also exist
\textit{figures of equilibrium with internal flows}:
\begin{itemize}
\itemsep=-1pt
\item[] the Dedekind ellipsoid (1861),

\item[] the Riemann ellipsoids (1861).
\end{itemize}

\smallskip\begin{small}{\bf Remark.}
The discovery of the Dedekind and Riemann ellipsoids was inspired by the
work of Dirichlet~\cite{ell36}, where the dynamical equations for a~liquid
homogeneous self-gravitating ellipsoid were obtained (for this system all
the above-mentioned figures of equilibrium are fixed points). For a~recent
review of dynamical aspects concerning liquid and gaseous self-gravitating
ellipsoids and a~detailed list of references, see~\cite{bmk}. We also note
the integrability cases found in a~related problem of gaseous
ellipsoids~\cite{gaffet}.

\end{small}\bigskip

While an enormous amount of research was devoted in the 19th and 20th
centuries to asymmetric figures of equilibrium (see, e.g., references
in~\cite{bmk,ell30}), the Maclaurin spheroid remains the most important
for applications to the theory of  the figures of planets. However, it is
well known that for all planets of the Solar System a~real compression is
different from the compression of the corresponding Maclaurin spheroid
obtained from the characteristics of the planet\footnote{Relevant
calculations can be easily performed using the formulae of
Section~\ref{sec2.3} and astronomic data available from the Internet.}.
Usually this difference is attributed to the density stratification of the
planet, which leads to the necessity of investigating inhomogeneous
figures of equilibrium.

For a~stratified fluid mass rotating as a~rigid body with small angular
velocity $\omega$, Clairaut\footnote{A.\,Clairot took part in the first
expeditions, which confirmed I.\,Newton's viewpoint that the Earth is
compressed from the poles.}\cite{clero} obtained the equation of
a~spheroid which is an equilibrium figure in the first order in
$\omega^2$. Subsequently investigations of such figures were continued in
the work of Laplace, Legendre and Lyapunov. Lyapunov obtained a~final
solution to this problem in the form of a~power series in the small
parameter $\omega^2$ and showed their convergence.

On the other hand, in~\cite{hamy,volterra}
and~\cite[Chapter~12]{pizzetti} it was shown that for a~stratified fluid
mass rotating as a~rigid body there exist no figures of equilibrium in the
class of ellipsoids. We present here in modern formulation a~theorem which
was proved in these works.\goodbreak

\smallskip
\textit{Suppose the body consists of a~self-gravitating, ideal,
stratified fluid. Assume that
\begin{itemize}
\item[--] the free surface of the fluid is an ellipsoid (it can be
    both three-axial and a~spheroid),
\item[--] the density distribution $\rho(\vec r)$ is such that
    the level surfaces $\rho(\vec r)={\rm const}$ are ellipsoids
    coaxial with the outer surface.
\end{itemize}
Then such a~fluid mass configuration cannot be the figure of equilibrium
rotating as a~rigid body about one of the principal axes.}
\smallskip

Hamy proved this theorem for the case of a~finite number of ellipsoidal
layers with constant density, Volterra generalized this result to the case
of continuous density distribution for a~homothetic stratification of
ellipsoids, and Pizzetti gave the simplest and most rigorous proof in the
general case for both continuous and piecewise constant density
distribution. Interestingly, there still appear papers (see,
e.g.,~\cite{kzs}) whose authors ``discover'' new ``solutions''
contradicting this theorem. Such publications show that there is still no
complete understanding regarding the equilibrium figures of celestial
bodies with stratified density. We also note that
A.\,Veronnet~\cite{veronnet} also tried to prove this theorem for the case
of continuous density distribution but made some errors.

If one admits the possibility that the angular velocity of fluid particles
is not constant for the entire fluid mass, then equilibrium figures for an
arbitrary axisymmetric form of the surface and density
stratification~\cite[Chapter~9]{pizzetti} are possible. For
example, in~\cite{chaplygin}\footnote{This work was not published during
the life-time of S.\,A.\,Chaplygin and appeared for the first time in his
posthumous collected works prepared by L.\,N.\,Sretenskii.}
S.\,A.\,Chaplygin explicitly showed a~spheroidal equilibrium figure with
a~nonuniform distribution of angular velocities for the case of homothetic
density stratification. It turns out that the surfaces with equal density
$\rho(\vec{r})={\rm const}$ do not coincide with the surfaces of equal
angular velocity $\omega(\vec{r})={\rm const}$. S.\,A.\,Chaplygin tried to
use the resulting solution to explain the dependence of the angular
velocity of rotation of the outer layers of the Sun on the latitude.

In~\cite{mms} an explicit solution of another kind was found for
which the equilibrium figure is a~spheroid consisting of two fluid masses
of different density $\rho_1\ne \rho_2$ separated by the spheroidal
boundary confocal to the outer surface, with each layer rotating at
constant angular velocity such that $\omega_1\ne \omega_2$. A
generalization of this solution to the case of an arbitrary finite number
of ``confocal layers'' was obtained in~\cite{esteban}.

In this paper we obtain a~generalization of this solution to the case of
an arbitrary confocal (both continuous and piecewise constant) density
stratification. For comparison, we also present Chaplygin's solution for
the homothetic\linebreak stratification. In addition, we show that in the
case of a~space with constant curvature the homogeneous (curvilinear)
spheroid is a~figure of equilibrium only under the condition of
a~nonuniform distribution of the angular velocities of fluid particles
$\omega(\vec{r})\ne {\rm const}$. In this case the solution can be
represented as a~power series in the space curvature.

\section{Equations of motion and axisymmetric equilibrium figures}

\subsection{Equations of motion in curvilinear coordinates}\label{subsec1_1}

In this case, to solve specific problems, it is convenient to use special
curvilinear (nonorthogonal) coordinates, which we denote by
${\vec q}=(q_1,q_2,q_3)$. Therefore, we first represent the equations
describing this system in an appropriate form.

Suppose that an element of the fluid has coordinates ${\vec q}$ at a~given
time~$t$. Let $\dot{\vec q}\,{=}\,(\dot q_1,\dot q_2,\dot q_3)$ denote the
rates of change of its coordinates during the motion. They depend on both
the coordinates ${\vec q}$ of the chosen element and time $t$: $\dot
q_i=\dot q_i({\vec q},t)$ and the total derivative of any function $f$ of
${\vec q}$, and $t$ is calculated from the formula
\begin{equation}
\label{eq_zv1}
\frac{df}{dt}=\frac{\partial f}{\partial t}+\sum\limits_i\frac{\partial f}{\partial q_i}\dot q_i.
\end{equation}
Let $\tens{G}=\|g_{ij}\|$ denote the metric tensor corresponding to these
coordinates. In the case of orthogonal coordinates
$\tens{G}={\rm diag}(h_1^2,h_2^2,h_3^2)$, where $h_i$~are the Lam\'{e} coordinates.

As is well known~\cite{kkp}, the equations of motion for a~fluid in a~potential field can be represented as
\begin{equation}
\label{eq_zv2}
\frac{d}{dt}\bigg(\frac{\partial T}{\partial \dot q_i}\bigg)-\frac{\partial T}{\partial q_i}=
    -\frac{\partial U}{\partial q_i}-\frac{1}{\rho}\frac{\partial p}{\partial q_i},
\end{equation}
where $\rho$ is the density, $p$~is the pressure, $U$~is the specific
potential of external forces, and $T$~is the specific kinetic energy of
the fluid calculated from the formula
$$
T=\frac12 \sum\limits_{i,j} g_{ij} \dot q_i\dot q_j.
$$

The continuity equations written in this notation become
\begin{equation}
\label{eq2_12}
\frac{\partial \rho}{\partial t}+\frac1{g}\sum\limits_i\frac{\partial}{\partial q_i}(\rho g\dot q_i)=0,\quad  g=\sqrt{\det \tens{G}}.
\end{equation}

In the case of a~self-gravitating fluid the gravitational potential
$U({\vec q},t)$ can be calculated from the equation
\begin{equation}
\label{eq_3}
\Delta U= 4\pi G\rho({\vec q},t),
\end{equation}
where  $G$~is the constant of gravitation and the Laplacian is given by the well-known relation
$$
\Delta=\frac1{g}\sum \frac{\partial}{\partial q_i}\bigg(g g^{ij}\,\frac{\partial}{\partial q_j}\bigg),\quad
\|g^{ij}\|=\tens{G}^{-1},
$$
assuming that outside the liquid body the density vanishes: $\rho=0$.

In the absence of external influences at the free boundary $\partial B$ of the fluid mass the pressure vanishes:
$$
p\big|_{\partial B}=0,
$$
and the gravitational potential and its normal derivative are continuous:
\begin{equation}
\label{eq4_12}
U_{\rm in}\big|_{\partial B}=U_{\rm out}\big|_{\partial B},\quad
\frac{\partial U_{\rm in}}{\partial n}\bigg|_{\partial B}= \frac{\partial U_{\rm out}}{\partial n}\bigg|_{\partial B},
\end{equation}
where the indices ${\rm in}$ and ${\rm out}$ denote the quantities inside
and outside the body, respectively.

\subsection{Steady-state axisymmetric flows}

To explore possible figures of equilibrium, we choose curvilinear
coordinates ${\vec q}=(r,\mu,\varphi)$, which are related to the Cartesian
coordinates as follows
$$
x=r\cos\varphi,\quad   y=r\sin\varphi,\quad  z=Z(r,\mu).
$$
Here the function $Z(r,\mu)$ is chosen so as to obtain a~free surface of
the fluid mass for one of the values $\mu=\mu_0$. Its specific form will
be defined by an appropriate problem statement.  The metric tensor is
given by
$$
\tens{G}=
\left(\begin{array}{ccc}
1+Z_r^2 & Z_r Z_{\mu} & 0\\
Z_rZ_{\mu} & Z_{\mu}^2 & 0\\
0 & 0 & r^2
\end{array}\right),\quad
g=\sqrt{\det\tens{G}}=rZ_{\mu},
$$
where $Z_r=\frac{\partial Z}{\partial r}$, $Z_{\mu}=\frac{\partial Z}{\partial\mu}$.

\smallskip\begin{small}{\bf Remark.}
P.\,Pizzetti~\cite{pizzetti} used usual cylindrical coordinates (i.e., he
set $\mu =z$), with the equation for the free surface being $F(x,z)=0$.
From a~practical point of view, this approach is inconvenient in searching
for specific equilibrium figures of a~stratified fluid.

\end{small}\smallskip

We shall seek a~steady-state solution of~(\ref{eq_zv2}), for which the
velocity distribution  has the form
\begin{equation}
\label{zvezdochka}
\dot r=0,\quad  \dot\mu=0,\quad  \dot\varphi=\omega(r,\mu),
\end{equation}
and the functions $U$, $p$, and $\rho$ do not depend on $\varphi$. Then,
substituting (\ref{zvezdochka}) into~(\ref{eq_zv2}) and (\ref{eq_3}), we obtain the
system of equations
\begin{equation}
\label{eq_zv}
\begin{array}{c}
\frac{\partial U}{\partial r}+\frac1{\rho}\,\frac{\partial p}{\partial r}=r\omega^2,\quad
\frac{\partial U}{\partial \mu}+\frac1{\rho}\,\frac{\partial p}{\partial \mu}=0,\\
\Delta_{r\mu} U=4\pi G\rho (r,\mu),\\
\Delta_{r\mu}=
\frac1{rZ_{\mu}}\,\frac{\partial}{\partial r}\Big(rZ_{\mu}\,\frac{\partial}{\partial r}\Big)+
    \frac1{Z_{\mu}}\,\frac{\partial}{\partial \mu}\Big(\frac{1+Z_{r}^2}{Z_{\mu}}\,\frac{\partial}{\partial \mu}\Big)
-\frac1{rZ_{\mu}}\Big(\frac{\partial}{\partial r}\Big(rZ_{r}\,\frac{\partial}{\partial \mu}\Big)+
\frac{\partial}{\partial \mu}\Big(rZ_{r}\,\frac{\partial}{\partial r}\Big)\Big),\\
p(r,\mu)\big|_{\mu=\mu_0}=0.
\end{array}
\end{equation}
Note that the continuity equation~(\ref{eq2_12}) holds identically.

We choose the function $Z(r,\mu)$ defining the curvilinear coordinates
in such a~way that all coordinate surfaces $\mu={\rm const}$ are compact,
and choose a~value of $\mu=\mu_0$ which corresponds to the boundary of the
fluid and defines the distribution of density $\rho(r,\mu)$. Then,
according to~(\ref{eq_zv}), after solving the equation for the potential
one can always choose a~distribution of pressure and of the squared
angular velocity, which satisfy the first pair of equations:
$$
\begin{array}{c}
p(r,\mu)=\int\limits_{\mu_0}^\mu \rho \frac{\partial U}{\partial \mu}\,d\mu,
\\
\omega^2(r,\mu)=\frac{1}{r\rho}\left( \rho_0\frac{\partial U}{\partial r}(r,\mu_0) +
\int\limits_{\mu_0}^\mu\left(\frac{\partial U}{\partial r}\frac{\partial \rho}{\partial \mu} -
\frac{\partial U}{\partial \mu}\frac{\partial \rho}{\partial r} \right) d\mu \right), \quad \rho_0=\rho(r,\mu_0).
\end{array}
$$
A possible obstruction to the existence of such equilibrium figures is
that $\omega^2(r,\mu)$, defined from these equations, may turn out to be
negative. The problem of equilibrium figures becomes more nontrivial
when we impose some restrictions on the distribution of angular velocity.

For example, L.\,Lichtenstein and R.\,Wavre (see~\cite{lihten}) found
sufficient conditions under which a~body obviously possesses a~plane of
symmetry.

\smallskip
\textbf{Theorem.}
\textit{Assume that for an inhomogeneous self-gravitating liquid body the
following is satisfied:}
\begin{enumerate}
\item[1.]  \textit{the fluid is at relative equilibrium where all particles
    rotate about the fixed axis $Oz$, and their angular velocity
    depends only on the distance to the axis of rotation:
    $\omega=\omega(r^2)$,}

\item[2.] \textit{the density is a~piecewise continuous function,}

\item[3.] \textit{the body consists of a~finite number of bounded regions
    whose boundaries have the topological type of a~sphere or a~torus.}
\end{enumerate}

\noindent \textit{Then the body possesses a~plane of symmetry perpendicular to the
axis $Oz$.}

\smallskip

It is also obvious that the center of mass lies on the intersection of the
symmetry plane with the axis of rotation $Oz$.

\section{Inhomogeneous figures with isodensity distribution of the angular velocity of layers}\label{sec2}

\subsection{General equations for  monotonic and piecewise constant density distribution}

We now consider the case where the level surfaces of stratification of
density $\rho$ coincide with the level surfaces of angular velocity
$\omega$ (i.e., the fluids of equal density move with equal angular
velocity); choosing them as coordinate lines $\mu={\rm const}$, we represent
this condition as
\begin{equation}
\label{eq6_12}
\rho=\rho(\mu),\quad  \omega=\omega(\mu).
\end{equation}
Eliminating the pressure from the first pair of equations of the
system~(\ref{eq_zv}) (multiplying them by~$\rho$ and differentiating the
first one with respect to $\mu$ and the second one with respect to $r$ and
subtracting  one from the other), we obtain
\begin{equation}
\label{eq_s1_}
\rho'(\mu)\,\frac{\partial U(r,\mu)}{\partial r}=r\big(\rho(\mu)\omega^2(\mu)\big)',
\end{equation}
where the prime denotes the derivative with respect to $\mu$.

{\bf 1.} We first consider the case where the density is nonconstant everywhere inside the body:
$$
\rho'(\mu)\not =0,\quad  \mbox{inside}.
$$

Then, according to~(\ref{eq_s1_}), the potential $U$ inside the body can be represented as
\begin{equation}
\label{eq7_12}
U(r,\mu)=\frac12u(\mu)r^2+v(\mu),
\end{equation}
and from the first pair of equations~(\ref{eq_zv}) we obtain the unknowns $p(r,\mu)$ and $\omega
(\mu)$ in the form
\begin{equation}
\begin{array}{c}
\label{biz7}
p=-\frac12P(\mu)r^2-Q(\mu),\quad  \omega^2(\mu)=u(\mu)-\frac{P(\mu)}{\rho(\mu)},\\
P(\mu)=\int\limits_{\mu_0}^{\mu} u'(\xi)\rho(\xi)\,d\xi,\quad
Q(\mu)=\int\limits_{\mu_0}^{\mu} v'(\xi)\rho(\xi)\,d\xi.
\end{array}
\end{equation}

Obviously,
$$
p(r,\mu)\Big|_{\mu=\mu_0}=0,\quad
\frac{d\omega^2}{d\mu}\bigg|_{\mu=\mu_0}\!\!=0.
$$

Hence, it follows that the figure of equilibrium of a~fluid with
stratification of density and angular velocity of the form~(\ref{eq6_12})
exists if and only if there exist functions $Z(r,\mu)$ and $u(\mu)$,
$v(\mu)$ satisfying the equation
\begin{equation}
\label{eq9_12}
\Delta_{r,\mu}\bigg(\frac12u(\mu)r^2+v(\mu)\bigg)=4\pi G\rho (\mu),
\end{equation}
and the potential inside the fluid mass has the form~(\ref{eq7_12}).

{\bf 2.} We now consider a~situation where in some layer the density takes
a~constant value:
$$
\rho(\mu)=\rho_0={\rm const},\quad  \mu\in(\mu_1, \mu_2),
$$
then, according to~(\ref{eq_s1_}), we conclude that the angular velocity
of the entire layer is also constant:
$$
\omega(\mu)=\omega_0={\rm const},\quad  \mu\in(\mu_1, \mu_2).
$$

Taking this into account, we integrate the first pair of
equations~(\ref{eq_zv}) and obtain the following relation for the function
$U+\frac{p}{\rho_0}$ in the layer:
\begin{equation}
\label{eq10_14}
U+\frac{p}{\rho_0}=\frac12 \omega_0^2r^2+\Phi_0,\quad  \Phi_0={\rm const}.
\end{equation}

\begin{wrapfigure}[14]{o}{35mm}
\includegraphics{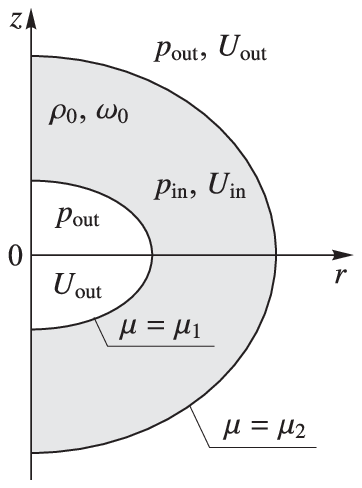}
\caption{}\label{fig2}
\end{wrapfigure}

Furthermore, at all points at the boundaries of the layer $\mu=\mu_i$,
$i=1,2$ (see Fig.~\ref{fig2}) the pressure inside and outside must be the same:
\begin{equation}
\label{eq11_12}
p_{\rm in}(r,\mu)\Big|_{\mu=\mu_i}=p_{\rm out}(r,\mu)\Big|_{\mu=\mu_i}\!.
\end{equation}

The potential in the layer also satisfies the Laplace equation
$$
\Delta_{r\mu}U_{\rm in}(r,\mu)=4\pi G\rho_0,
$$
and at the boundaries the conditions~(\ref{eq4_12}) hold.

\subsection{The family of confocal spheroids}\label{subsec2.2}

Consider a~particular case in which the sought-for solution exists. We
shall show that in the case of confocal stratification of the density of a~spheroid the gravitational potential is written as~(\ref{eq7_12}).

Choose the parameterization of confocal stratification in ${\mathbb R}^3$ as
follows
$$
\frac{x^2+y^2}{d^2(1+\mu^2)}+\frac{z^2}{d^2\mu^2}=1,\quad  \mu\in[0,+\infty),
$$
\begin{wrapfigure}[15]{o}{40mm}
\includegraphics{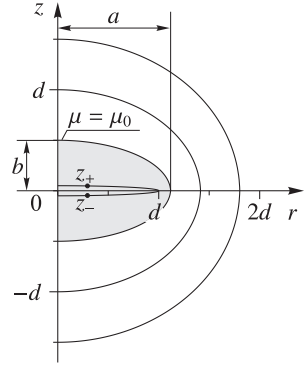}
\caption{Meridional sections of the surfaces $\mu={\rm const}$}\label{fig1}
\end{wrapfigure}
where $d$~is the focal distance of the meridianal section (see
Fig.~\ref{fig1}). Thus, the parameter $\mu$ defines the ratio between the
small semi-axis of the spheroid and the focal distance, and the
eccentricity $e$ is expressed by the formula
\begin{equation}
\label{eq_s0}
e=\frac1{\sqrt{1+\mu^2}}.
\end{equation}
Expressing $z$, we find
\begin{equation}
\label{biz10}
Z(r,\mu) =\pm\sqrt{d^2\mu^2-r^2\frac{\mu^2}{\mu^2+1}}.
\end{equation}
If the boundary of the spheroid filled with a~fluid has semi-axes $a$ and $b$ (see
Fig.~\ref{fig1}), then the focal distance $d$ and the coordinate of the boundary $\mu_0$
are defined by
\begin{equation}
\label{eq10_12_}
d=\sqrt{\displaystyle a^2-b^2},\quad  \mu_0=\frac{b}{\sqrt{\displaystyle a^2-b^2}}.
\end{equation}

\smallskip\begin{small}{\bf Remark.}
It can be shown that for a~prolate spheroidal stratification \Big(i.\,e.,
for $\frac{r^2}{d^2\mu^2}+\frac{z^2}{d^2(\mu^2+1)}\,{=}\,1$\Big) this solution leads
to a~negative square of the angular velocity of rotation of the layers
($\omega^2(\mu)<0$), therefore, we will not consider it.

\end{small}\smallskip

\begin{pro}
\label{pbiz1}
The gravitational potential for a~spheroid with confocal\linebreak
stratification has the form
\begin{equation}
\label{biz4}
U=\frac{k}{2}\left(\frac{1}{2}\frac{r^2\widetilde  u(\mu)}{1+\mu^2} + d^2 \widetilde  v(\mu)\right), \ k=4\pi G.
\end{equation}
For the internal points
\begin{equation}
\label{biz2}
\begin{gathered}
\widetilde {u}^{\rm in}  =I_0(\mu)((1+3\mu^2)\arcctg(\mu) - 3 \mu) - I_1(\mu)(1+3\mu^2) \\
\widetilde {v}^{\rm in} =-I_0(\mu)((1+\mu^2)\arcctg(\mu) -\mu) +I_1(\mu)(1+\mu^2) + 2I_2(\mu)
\end{gathered}
\end{equation}
$$
\begin{gathered}
I_0(\mu)=\int\limits_0^{\mu}\rho(\xi)(1+3 \xi^2)\,d \xi, \quad
I_1(\mu)=\int\limits_{\mu_0}^{\mu}\rho(\xi)((1+3\xi^2)\arcctg(\xi) - 3\xi) \,d \xi, \\
I_2(\mu)=\int\limits_{\mu_0}^{\mu}\xi\rho(\xi) \,d\xi.
\end{gathered}
$$
For the external points
\begin{equation}
\label{biz3}
\widetilde {u}^{\rm out}=I_0(\mu_0)((1+3\mu^2)\arcctg(\mu) - 3 \mu), \quad \widetilde {v}^{\rm out}=I_0(\mu_0)(\mu - (1+\mu^2)\arcctg(\mu)).
\end{equation}
\end{pro}

{\small
\proof We shall search for a~potential in the form (\ref{biz4}).
Then Eq. (\ref{eq9_12}) leads to two linear equations for the functions $\widetilde  u(\mu)$ and $\widetilde  v(\mu)$:
\begin{equation}
\label{eq10_12}
\frac{d}{d\mu}\bigg((1+\mu^2)\frac{d \widetilde  u}{d\mu}\bigg)-6\widetilde  u+4 \rho(\mu)=0,\\
\frac{d}{d\mu}\bigg((1+\mu^2)\frac{d\widetilde  v}{d\mu}\bigg)+2\widetilde  u-2(1+\mu^2) \rho(\mu)=0,
\end{equation}
As is well known, the solution~(\ref{eq10_12}) is represented as the superposition
\begin{equation}
\label{eq_s4}
\widetilde  u(\mu)=\widetilde  u_0(\mu)+\widetilde  u_p(\mu),\quad
\widetilde  v(\mu)=\widetilde  v_0(\mu)+\widetilde  v_p(\mu),
\end{equation}
where $\widetilde  u_0$ and $\widetilde  v_0$ are a~general solution of the
homogeneous system
(when $\rho(\mu)=0$), while $\widetilde  u_p$ and $\widetilde  v_p$ are a~particular solution
of the inhomogeneous system. In this case one can choose
\begin{equation}
\label{eq_s5}
\begin{gathered}
\widetilde  u_0(\mu)=A_1(1+3\mu^2)+A_2\left((1+3\mu^2)\arcctg\mu -3\mu\right),\\
\widetilde  v_0(\mu)=-A_1\mu^2+A_2\left((1-\mu^2)\arcctg\mu +\mu\right)
    +A_3\arcctg\mu+A_4.
\end{gathered}
\end{equation}
Using a~modification of the method of variation of constants, the
particular solution can be represented as single integrals:
\begin{equation}
\label{eq_s6}
\begin{gathered}
\begin{aligned}
\widetilde  u_p(\mu)&= \Big((1+3\mu^2)\arcctg\mu-3\mu\Big)\times\\
&\times\int\limits^{\mu}_{\mu_s}(1+3\xi^2)\rho(\xi)\,d\xi-
    (1+3\mu^2)\int\limits_{\mu_s}^{\mu} \bigg((1+3\xi^2)\arcctg\xi -3\xi\bigg)\rho(\xi)\,d\xi,
\end{aligned}\\
\widetilde  v_p(\mu)=\int\limits_{\mu_s}^{\mu}(\arcctg\xi )S(\xi)\,d\xi -\arcctg\mu \int\limits_{\mu_s}^{\mu}S(\xi)\,d\xi ,\\
2S(\mu)=(\mu^2+1)\rho(\mu)-\widetilde  u_p(\mu),
\end{gathered}
\end{equation}
In the general case, for each of the integrals an arbitrary constant  can
be chosen as the lower bound $\mu_s$ in~(\ref{eq_s6}).

The conditions which must be satisfied by the potential have the form

{\bf 1}. Away from the spheroid, the potential must tend to zero:
\begin{equation}
\label{eq_s2}
\lim\limits_{\mu\to\infty} \frac{\widetilde  u^{\rm out}(\mu)}{1+\mu^2}=0,\quad
\lim\limits_{\mu\to\infty} \widetilde  v^{\rm out}(\mu)=0.
\end{equation}

{\bf 2}. At the boundary of the spheroid $\mu=\mu_0$, the potential must be a~smooth function:
\begin{equation}
\label{eq_s1}
\begin{gathered}
\widetilde  u^{\rm in}(\mu_0)=\widetilde  u^{\rm out}(\mu_0), \quad  \widetilde  v^{\rm in}(\mu_0)=\widetilde  v^{\rm out}(\mu_0), \\
\widetilde  u'^{\rm in}(\mu_0)=\widetilde  u'^{\rm out}(\mu_0), \quad  \widetilde  v'^{\rm in}(\mu_0)=\widetilde  v'^{\rm out}(\mu_0).
\end{gathered}
\end{equation}

{\bf 3}. As $\mu\to 0$, the potential on the section $z=0$, $r\in(0,d)$
must be a~smooth function, i.e., the values of its derivatives must be the
same at the points $z_+$ and $z_-$ as $\mu\to 0$ (see Fig.~\ref{fig2}).
This yields the condition
\begin{equation}
\label{eq_s3}
\widetilde  u'_{\rm in}\Big|_{\mu=0}=0,\quad
\widetilde  v'_{\rm in}\Big|_{\mu=0}=0.
\end{equation}

Let us satisfy the first condition (\ref{eq_s2}). To do so, we express the
potential outside as a~power series in $\frac{1}{\mu}$:
$$
\frac{\widetilde  u^{\rm out}(\mu)}{1+\mu^2}=3A_1^{\rm out} + O\bigg(\frac1{\mu}\bigg), \quad
\widetilde  v^{\rm out}=-A_1^{\rm out}\mu^2 +A_4^{\rm out} + O\bigg(\frac1{\mu}\bigg)
$$
to give $A_1^{\rm out}=A_4^{\rm out}=0$.

Next, we satisfy the condition (\ref{eq_s1}). To simplify the system
(\ref{eq_s2}), we choose a~particular solution in such a~way that it
vanishes on the surface. It is easily seen that this can be achieved by
choosing $\mu_s=\mu_0$. Moreover, in this case Eqs. (\ref{eq_s1}) are
satisfied if we set $A_1^{\rm in}=A_4^{\rm in}=0$, $A_2^{\rm out}=A_2^{\rm
in}$, $A_3^{\rm out}=A_3^{\rm in}$.

From Eqs. (\ref{eq_s3}) we find two remaining constants $A_2^{\rm in}$ and $A_3^{\rm in}$:
$$
A_2^{\rm in}=\int\limits_0^{\mu_0} (1+3\xi^2)\rho(\xi)\,d\xi,\quad
A_3^{\rm in}=-2\int\limits_0^{\mu_0} S(\xi)\,d\xi.
$$

Now, in order to obtain the relations (\ref{biz2}), we only need to simplify the expression for $A_3^{\rm in}$:
$$
\begin{gathered}
A_3^{\rm in}=-2\int\limits_0^{\mu_0}(1+\mu^2)\rho(\mu)\,d\mu+2\int\limits_0^{\mu_0} \widetilde  u_p(\mu)\,d\mu,\\
2\widetilde  u_p(\mu)=2\left( \psi_1(\mu) \int\limits_{\mu_0}^{\mu}\psi_2(\xi)\rho(\xi)\,d\xi-
    \psi_2(\mu) \int\limits_{\mu_0}^{\mu}\psi_1(\xi)\rho(\xi)\,d\xi \right)\!,\\
\psi_1(\mu)=(1+3\mu^2)\arcctg\mu-3\mu,\quad
\psi_2(\mu)=1+3\mu^2.
\end{gathered}
$$
We take the second integral in the expression for $A_3^{\rm in}$ by parts. To do this, we define the primitives
$$
\begin{gathered}
\Psi_1(\mu)=\mu\big((1+\mu^2)\arcctg\mu-\mu\big),\quad  \Psi'_1(\mu)=\psi_1(\mu),\\
\Psi_2(\mu)=\mu(\mu+1),\quad  \Psi'_2(\mu)=\psi_2(\mu),
\end{gathered}
$$
to give
$$
\int\limits_0^{\mu_0} 2\widetilde  u_p(\mu)\,d\mu=\int\limits_0^{\mu_0} \big(\psi_1(\mu)\Psi_2(\mu)-\psi_2(\mu)\Psi_1(\mu)\big)\rho(\mu)\,d\mu=
    -2\int\limits_0^{\mu_0} \mu^2\rho(\mu)\,d\mu.
$$
Thus, we finally obtain
$$
A_2^{\rm in}=-2\int\limits_0^{\mu_0} (1+3\xi^2)\rho(\xi)\,d\xi.\quad
$$

Writing the solution (\ref{eq_s6}) with the known integration constants,
taking the iterated integrals in $\widetilde  v_p(\mu)$ by parts, as was done
above, and reducing similar terms, we obtain (\ref{biz2}) and
(\ref{biz3}). \qed}

\smallskip\begin{small}{\bf Remark.}
If we make a~change of the variable $\mu=i x$ in Eqs. (\ref{eq10_12}),
they take the form of inhomogeneous Legendre equations with $n=2$ and
$n=1$.

\end{small}\smallskip

As a~consequence of this representation of the potential, we obtain the
well-known Maclaurin theorem~\cite{ell30} in the case of a~spheroid.

\begin{teo}
The gravitational potential that is produced by an inhomogeneous spheroid
with confocal stratification and density $\rho(\mu)$ is at the external
point the same as the potential of a~homogeneous spheroid with the density
\[
\langle  \rho \rangle =\frac{1}{\mu_0(1+\mu_0^2)}\int\limits_0^{\mu_0} (1+3\xi^2)\rho(\xi)\,d\xi.
\]
\end{teo}

According to Proposition \ref{pbiz1}, the family of confocal spheroids
satisfies the condition (\ref{eq7_12}), and hence the level surfaces of
angular velocity are also confocal spheroids. After integrating $P(\mu)$
by parts the final expression for the angular velocity of the layers can
be represented as
\begin{equation}
\label{biz5}
\begin{gathered}
\frac{\omega(\mu)^2}{2 \pi G}=I_0(\mu_0)\frac{\rho(\mu_0)}{\rho(\mu)}\frac{(1+3\mu^2_0)\arcctg(\mu_0) - 3\mu_0}{1+\mu^2_0} -{} \\
- \frac{2}{\rho(\mu)}\int\limits_{\mu}^{\mu_0}\rho'(\xi)\frac{I_0(\xi)((1+3\xi^2)\arcctg(\xi) -3\xi ) - I_1(\xi)(1+3\xi^2)}{1+\xi^2}\,d \xi.
\end{gathered}
\end{equation}

From this relation, setting $\mu=\mu_0$, we obtain the following result:

\begin{teo}
For an arbitrary confocal stratification the angular velocity on the outer
surface of the inhomogeneous spheroid is the same as the angular velocity
of the Maclaurin spheroid with density $\langle \rho\rangle $:
\begin{equation}
\label{biz6}
\frac{\omega^2_0}{2\pi G \langle \rho\rangle }=\mu_0((1+3\mu_0^2)\arcctg(\mu_0) -3 \mu_0),
\end{equation}
where $\langle \rho\rangle $ is the average density of the spheroid.
\end{teo}

\subsection{The homogeneous Maclaurin spheroid}\label{sec2.3}

Let the density be constant everywhere inside some spheroid:
$$
\rho(\mu)=\left\{
\begin{aligned}
&0,&\quad &\mu_0<\mu,\\
&\rho_0,&&0<\mu\le\mu_0,
\end{aligned}
\right.
$$
where $\mu_0$~is defined by~(\ref{eq10_12_}). In this case we find the
gravitational potential from Proposition \ref{pbiz1}.  Inside the spheroid
it can be represented as
$$
\begin{gathered}
U=2\pi G\left(\frac{1}{2}\frac{r^2\widetilde  u^{\rm in}(\mu) }{1 + \mu^2} + d^2\widetilde  v^{\rm in}(\mu)\right),
\\
u^{\rm in}(\mu)=\rho_0\big(\mu_0(1+3\mu^2)((1+\mu_0^2)\arcctg \mu_0 - \mu_0)-2\mu^2\big),
\\
v^{\rm in}(\mu)=\rho_0(1+\mu_0^2)\big(\mu^2-\mu_0(1+\mu^2)\arcctg \mu_0 \big).
\end{gathered}
$$

Next, from (\ref{biz5}) and taking into account the relationship
(\ref{eq_s0}) between $\mu_0$ and the eccentricity, we obtain the
well-known expression for the angular velocity of the Maclaurin spheroid
$$
\frac{\omega_0^2}{2\pi G\rho_0}=\mu_0\Big( (1+3\mu_0^2)\arcctg\mu_0-3\mu_0\Big)=
\frac{\sqrt{1-e^2}}{e^3}\Big((3-2e^2)\arcsin e-3e\sqrt{1-e^2}\Big).
$$

Using (\ref{biz7}), we find the pressure for the Maclaurin spheroid:
\begin{equation}
\label{biz8}
\frac{p}{2 \pi G \rho_0^2}=\frac{(\mu_0^2 - \mu^2)(1 - \mu_0\arcctg\mu_0)}{1+\mu^2} (d^2(1+\mu^2)(1+\mu_0^2) - r^2).
\end{equation}
It can be shown that the level surfaces (\ref{biz8}) are homothetic
spheroids. To do so, we use a~relation defining the homothetic
stratification, which in our case takes the form
$$
\frac{r^2}{d^2(1+\mu_0^2)}+\frac{z^2}{d^2\mu_0^2}=m,
$$
and (\ref{biz10}), we find
$$
r=\frac{d^2(1+\mu_0^2)(1+\mu^2)(m\mu_0^2-\mu^2)}{\mu^2-\mu^2_0}.
$$
Then, substituting $r$ into (\ref{biz8}), we obtain
$$
\frac{p}{2 \pi G \rho_0^2}=d^2\mu_0^2(1+\mu_0^2)(1 - \mu_0\arcctg\mu_0) (1 - m).
$$

\subsection{A spheroid with piecewise constant density distribution}

We now consider a~spheroid with piecewise constant density, i.e.,
consisting of a~sequence of embedded homogeneous layers with different
densities. We will number the outer layer, as before, by the index $0$ and
the last internal layer by the index $n$. Thus, we obtain a~spheroid
consisting of $n+1$ layers:
$$
\rho(\mu)=\left\{
\begin{aligned}
&0,     &\quad  &\mu_0<\mu,\\
&\rho_0,&\quad  &\mu_1<\mu<\mu_0,\\
&\rho_1,&\quad  &\mu_2<\mu<\mu_1, \\
& \ldots  ,&\quad  &  \quad \quad  \ldots \\
&\rho_n,&\quad  &0<\mu<\mu_n.
\end{aligned}
\right.
$$

The case of two layers of different density (in our notation $n=1$) is
considered in~\cite{mms}, and the generalization of this case to an
arbitrary number of layers is found in~\cite{esteban}. Interestingly,
almost all calculations presented below are contained in~\cite{hamy},
although he used them not to search for new figures of equilibrium but to
prove the absence of inhomogeneous figures of equilibrium with rigid body
rotation (see the Introduction).

From~(\ref{eq10_14}) we find that the pressure inside the $k$-th layer is given by
$$
\frac{p^{(k)}}{\rho_{i}}=\pi G r^2\bigg( \frac{\omega^2_{k}}{2\pi G}-\frac{\widetilde  u_{\rm in}(\mu)}{1+\mu^2} \bigg)+
   2 \pi Gd^2\widetilde  v_{\rm in}(\mu)+\Phi_{k},\quad  k=0,1,\ldots,n.
$$
where $\mu_k<\mu<\mu_{k+1}$.

Further, taking into account that the pressure at the outer boundary is
zero and the potential and the pressure at the boundary between the layers
change continuously, we obtain the following relations for unknown angular
velocities:
$$
\begin{gathered}
\frac{\Delta_0\omega_0^2}{2 \pi G}=\Delta_0\frac{\widetilde  u_{\rm in}(\mu_0)}{1 +\mu_0^2},
\\
\ldots,
\\
\frac{\rho_n \omega_n^2}{2 \pi G}=\frac{\rho_{n-1}\omega_{n-1}^2}{2 \pi G} + \Delta_n\frac{\widetilde  u_{\rm in}(\mu_n)}{1 +\mu_n^2},
\\
\Delta_0=\rho_0, \ \Delta_1=\rho_1-\rho_0, \ \ldots \ , \Delta_n=\rho_n-\rho_{n-1}.
\end{gathered}
$$
This yields the angular velocity for the $k$-th layer in the form
$$
\frac{\rho_k\omega_k^2}{2 \pi G}=\sum \limits_{i=0}^{k} \Delta_i\frac{\widetilde  u_{\rm in}(\mu_i)}{1 +\mu_i^2}.
$$

We obtain the expression for $\widetilde  u_{\rm in}(\mu_i)$ from (\ref{biz2}):
$$
\widetilde  u_{\rm in}(\mu_i)=I_0(\mu_i)((1+3\mu_i^2)\arcctg\mu_i - 3\mu_i) - I_1(\mu_i)(1+3\mu_i^2).
$$
To calculate $I_0(\mu_i)$ and $I_1(\mu_i)$, we use the Heaviside function
$$
\theta(x)=\left\{
\begin{aligned}
&0,&\quad  &x<0,\\
&1,& &x\geq 0,
\end{aligned}
\right.
$$
and represent the density of the spheroid under consideration as
$$
\rho(\mu)=\sum \limits_{i=0}^{n}\Delta_i\theta(\mu_i - \mu).
$$
Integrating, we find that
$$
\begin{gathered}
I_0(\mu_i)=\sum \limits_{j=i+1}^{n}\Delta_j\mu_j(1+\mu_j^2)
\\
I_1(\mu_i)=\sum \limits_{i=0}^{j}\Delta_j\left(\frac{2\mu_i^2}{1+3\mu_i^2} - \mu_j((1+\mu_j^2)\arcctg \mu_j - \mu_j)\right).
\end{gathered}
$$
As a~result, we obtain an expression for the angular velocity of the $k$-th layer in the form
\begin{equation}
\label{biz15}
\begin{gathered}
\frac{\rho_k\omega_k^2}{2\pi G}=\sum \limits_{i=0}^{k}\Delta_i
    \left(\frac{1+3\mu_i^2}{1+\mu_i^2} \sum \limits_{j=0}^{i}\Delta_j
    \left(\mu_j((1+\mu_j^2)\arcctg\mu_j - \mu_j) - \frac{2\mu_i^2}{1+3\mu_i^2} \right) \right.+ \\
+\left.\frac{(1+3\mu_i^2)\arcctg\mu_i - 3\mu_i}{1+\mu_i^2}\sum \limits_{j=i+1}^{n}\Delta_j\mu_j(1+\mu_j^2)    \right).
\end{gathered}
\end{equation}

\subsection{A spheroid with continuous density distribution}

To keep track of the dependence of the angular velocity of the layers on
the change in density, we consider an inhomogeneous spheroid with
different functions of density distribution of the following form:
\begin{equation}
\label{biz12}
\rho(\mu)=\rho_{n}^{(0)}(1 - \alpha_n \mu^n), \ n=2,4,6,
\end{equation}
where $\rho_{n}^{(0)}$ and $\alpha_n$ are some constants (note that
$\rho_{n}^{(0)}$ has the meaning of density at the center of the
spheroid). We will determine their values from
the given average density of the body
$
\langle \rho\rangle =\frac{\int \rho dV}{\int dV}
$
and the given ratio between the density on the surface and the average density of the
body $\varepsilon=\frac{\langle \rho\rangle }{\rho(\mu_0)},$
$$
\begin{gathered}
\alpha=\frac{(1+n)(3+n)(1+\mu_0^2)(1-\varepsilon)\mu_0^{-n}}{(3+n)(1-\varepsilon (1+n)(1+\mu_0^2)) + 3(1+n)\mu_0^2}
\\
\rho_0=\langle  \rho \rangle \frac{(3+n)(\varepsilon(1+n)(1+\mu_0^2)-1) - 3(1+n)\mu_0^2}{n\varepsilon((1+n)\mu_0^2+3+n)}.
\end{gathered}
$$

As an example, assume that the eccentricity $e_0$ and $\varepsilon$, which are
the same as the data of the Earth~\cite{david}:
$$
e_0=0.08181, \quad  \varepsilon=2.5.
$$

\begin{figure}[!ht]
\parbox[t]{0.48\textwidth}{\includegraphics[scale=0.95]{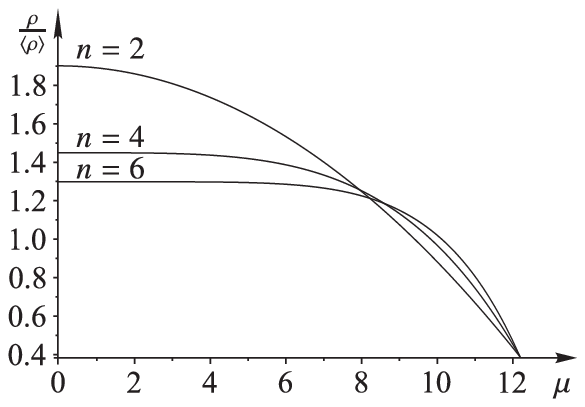}} \hfill
\parbox[t]{0.48\textwidth}{\includegraphics[scale=0.95]{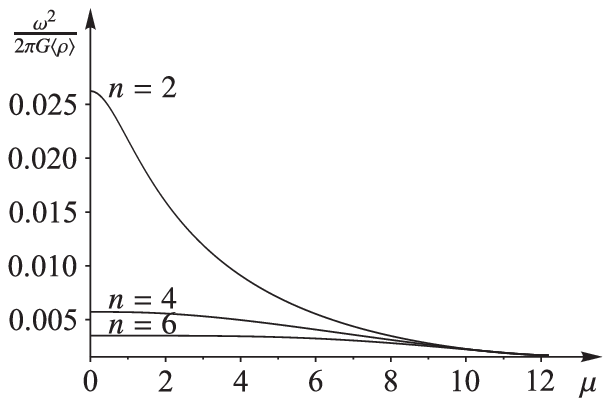}} \\
\parbox[t]{0.48\textwidth}{\caption{A graph showing the dependence of the relation $\frac{\rho}{\langle \rho\rangle }$ on the layer $\mu$}\label{f1}}\quad
\parbox[t]{0.48\textwidth}{\caption{A graph showing the dependence of the angular velocity on the layer $\mu$}\label{f2}}
\end{figure}

Figure~\ref{f1} shows the dependences of $\frac{\rho}{\langle \rho\rangle
}$ on the coordinate of the layer $\mu$ for are spheroid with density
distribution described by (\ref{biz12}). As we can see, the density
increases most sharply at the center of the spheroid for $n=2$ and then,
as $n$ increases, the density decreases.

To find the angular velocity, we substitute the density distributions
(\ref{biz12}) into (\ref{eq11_12}) and obtain the dependence of the
angular velocity on the layer. A graph of this dependence is shown in Fig.~\ref{f2}. (Since the
explicit formulae for $\omega(\mu)$ are unwieldy, we do not present them
here.)

For the angular velocity with density distribution (\ref{biz12}) one may draw the
following conclusion from Fig.~\ref{f2}: {\it the closer the center of the
spheroid, the larger the angular velocity; specifically, the larger the
value of density at the center of the spheroid (with $n=2$), the larger
the increase in the angular velocity.}

\begin{figure}[!ht]
\centering\includegraphics{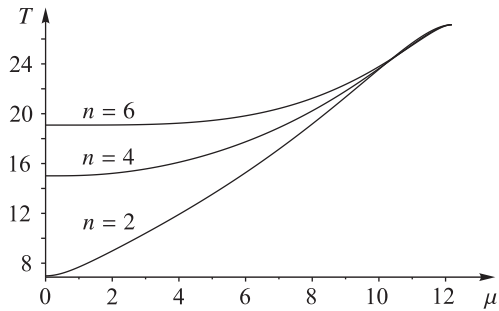}
\caption{The period of revolution $T$ depending on the layer $\mu$}
\label{f3}
\end{figure}

Next, we calculate the numerical value of the dependence of the period of
revolution for each layer. If we assume the average density to be the same
as that of the Earth $\langle \rho\rangle =5.51$g/cm$^3$, then we obtain the
dependences of $T(\mu)$ presented in Fig.~\ref{f3}.

\section{The Chaplygin problem~---\\ a~spheroid with homothetic density distribution}\label{sec3}

\begin{figure}[!b]
\centering
\includegraphics{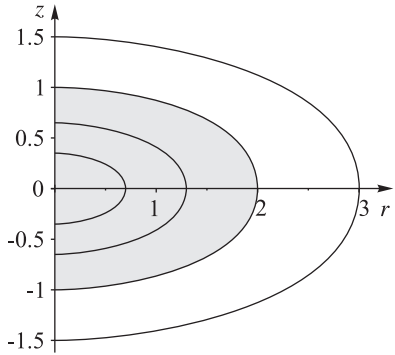}
\smallskip
\caption{Meridional sections of the surfaces $\sigma={\rm const}$ with
homothetic stratification}\label{fign}
\end{figure}

As is well known, the homothetic stratification is given by
$$
\frac{z^2}{b^2}+\frac{r^2}{a^2}=\sigma,\quad  \sigma \in[0,+\infty),
$$
where, assuming that $a$ and $b$~are the principal semi-axes of a~spheroid filled with a~fluid (see Fig.~\ref{fign}), we obtain
$$
\sigma_0=1,\quad  Z(r,\sigma)=\pm b\sqrt{\sigma-\frac{r^2}{a^2}}.
$$
Again we set
$$
\rho=\left\{
\begin{array}{lll}
\rho(\sigma) \mbox{ (does not depend on $r$)}, &\sigma\le 1,\\
0, &\sigma> 1.\\
\end{array}
\right.
$$
Using the second of Eqs.~(\ref{eq_zv}) and noting that
$p\big|_{\sigma=1}=0$, we obtain the pressure, which can be represented as
$$
p(r,\sigma)=\rho_1  U(r,1)-\rho(\sigma) U(r,\sigma)+\int\limits_{1}^{\sigma} U(r,\sigma)\,\frac{\partial\rho}{\partial\sigma}\,d\sigma,\quad
    \rho_1=\rho(1).
$$

In a~similar manner, substituting the pressure from the first of Eqs.~(\ref{eq_zv}),
we obtain
\begin{equation}
\label{biz17}
\omega^2(r,\sigma)=\frac{1}{r\rho(\sigma)}\bigg(\rho_1\frac{\partial U}{\partial r}(r,1)+
\int\limits_{1}^{\sigma} \frac{\partial U}{\partial r}(r,\sigma)\frac{\partial\rho}{\partial\sigma}\,d\sigma\bigg).
\end{equation}
Thus, to complete the solution, we only need to find the potential from the equation
$$
\Delta_{r,\sigma} U(r,\sigma)=4\pi G\rho(\sigma).
$$

In~\cite{ferrers} a~convenient integral representation of the potential
for a~(three-axial) ellipsoid with homothetic density stratification is
obtained. Applying it to the case of the spheroid $\sigma=1$ gives
\begin{equation}
\label{biz16}
\begin{gathered}
U^{\rm in}(r,z)=\pi Ga^2b^2\int\limits_0^{\infty} \frac{f(1)-f\left(\frac{r^2}{a^2+s}+\frac{z^2}{b^2+s}\right)}{\Delta(s)}\,ds,\\
U^{\rm out}(r,z)=\pi Ga^2b^2\int\limits_{s_0}^{\infty} \frac{f(1)-f\left(\frac{r^2}{a^2+s}+\frac{z^2}{b^2+s}\right)}{\Delta(s)}\,ds,\\
\Delta(s)=(a^2+s)\sqrt{b^2+s},
\end{gathered}
\end{equation}
where the function $f(\sigma)$ is related with the density of the fluid by
$$
\rho(\sigma)=\frac{df(\sigma)}{d\sigma},
$$
and the quantity $s_0$ for given $(r,z)$, which correspond to a~point
outside the liquid spheroid, is defined as the root of the equation
$$
\frac{r^2}{a^2+s_0}+\frac{z^2}{b^2+s_0}=1.
$$

\begin{figure}[!ht]
\centering\includegraphics[scale=0.95]{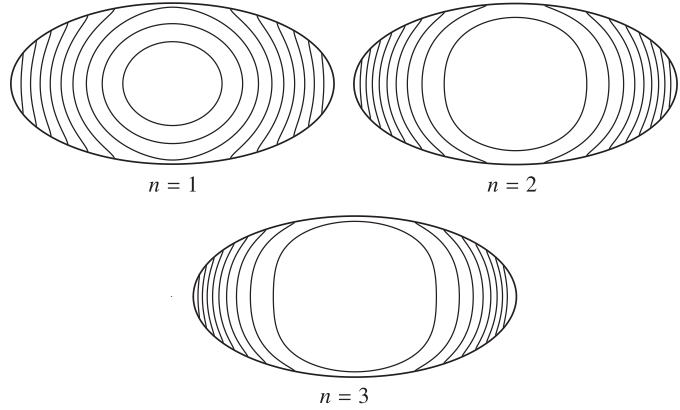}
\caption{Meridional sections of the surfaces $\frac{\omega^2}{2\pi G \langle \rho\rangle }={\rm const}$ with equal spacings}
\label{f3_}
\end{figure}

\begin{figure}[!ht]
\centering\includegraphics[scale=0.95]{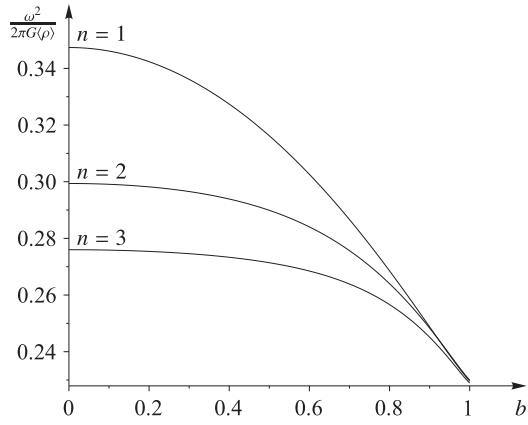}
\caption{The change of $\frac{\omega^2}{2\pi G\langle \rho\rangle }$ along the small semi-axis $b$ for different $n$}
\label{fig8}
\end{figure}

As an example, we consider the density distribution of the form
\begin{equation}
\label{biz18}
\rho(\sigma)=\rho_0(1 - \alpha\sigma^n), \quad  n=1,2,3.
\end{equation}
Given the average density $\langle \rho\rangle $ of the body and the ratio between the
densities at the center and on the surface $\eta=\frac{\rho_0}{\rho_1}$,
we now define the constants $\rho_0$ and $\alpha$:
\begin{equation}
\label{biz19}
\alpha=\frac{\eta - 1}{\eta}, \quad  \rho_0=\frac{\eta(3 + 2n)\langle \rho \rangle }{3 + 2n\eta}.
\end{equation}
Set
$$
\eta=5, \quad  \frac{b}{a}=\frac{1}{2}.
$$
Further, we find the potential from (\ref{biz16}) and obtain the angular
velocity from (\ref{biz17}). The meridional sections of the surfaces
$\frac{\omega^2}{2 \pi G \langle  \rho \rangle }={\rm const}$ with equal spacings for
different $n=1,2,3$ are shown in Fig.~\ref{f3_}. The graphs of change in
the relation $\frac{\omega^2}{2\pi G\langle \rho\rangle }$ along the small semi-axis $b$ is shown
in Fig.~\ref{fig8}.

For the densities from Figs.~\ref{f3_} and~\ref{fig8} one can draw the following conclusions:

1. {\it The closer the center of the spheroid, the slower the change in the angular velocity.}

2. {\it For $n=1$ the level surfaces near the center of the spheroid are
concentric spheres. Further, as $n$ increases, the region in which the
level lines are closed surfaces increases. For $n>1$ these closed surfaces
are no longer surfaces of the second order.}

Let us consider in more detail the angular velocity at the boundary of the
spheroid at the equator with densities of the form (\ref{biz18}), but now
with an arbitrary $n$. From (\ref{biz17}), changing the variable
$s=a^2(t-1)$, we obtain the angular velocity on the surface:
$$
\begin{aligned}
\frac{\omega^2_n(r,1)}{2 \pi G}&=\rho_0e^2\sqrt{1-e^2}\int_1^{\infty}\frac{t-1}{t^2(t-e^2)^{3/2}}\times\\
&\times
    \left(1 - \frac{\alpha t^{-n}}{(t-e^2)^n}\left((t-1)e^2\frac{r^2}{a^2} + t(1-e^2)\right)^n\right)dt,
\end{aligned}
$$
that is, for $r=a$ we have
$$
\frac{\omega^2_n(a,1)}{2 \pi G}=\rho_0e^2\sqrt{1-e^2}\int_1^{\infty}\frac{(t-1)(1-\alpha t^{-n})}{t^2(t-e^2)^{3/2}}dt.
$$
Explicitly integrating gives
\begin{align}
\label{biz19_}
\frac{\omega^2_n(a,1)}{2 \pi G}&=\rho_0\omega_m^2 + \frac{2 \alpha \rho_0e^2}{3+2n}\times\\
&\times\left(\frac{\sqrt{1-e^2)}(2n(1-e^2) +3 -2e^2)}{5+2n}
F\left(\frac{3}{2},n+\frac{5}{2}, n+\frac{7}{2},e^2\right) - 1\right),\notag
\end{align}
where $\omega_m^2$ is the dimensionless angular velocity of the Maclaurin spheroid:
$$
\omega_m^2=\frac{\sqrt{1-e^2}}{e^3}\Big((3-2e^2)\arcsin e-3e\sqrt{1-e^2}\Big).
$$
Substituting the expression (\ref{biz19_}) into the relation for the angular velocity, we obtain for two values of $n$
$$
\frac{\omega^2_0(a,1)}{2 \pi G\rho_0(1-\alpha)}=\frac{\omega^2_{\infty}(a,1)}{2 \pi G \rho_0}=\omega_m^2.
$$

\begin{figure}[!ht]
\centering\includegraphics{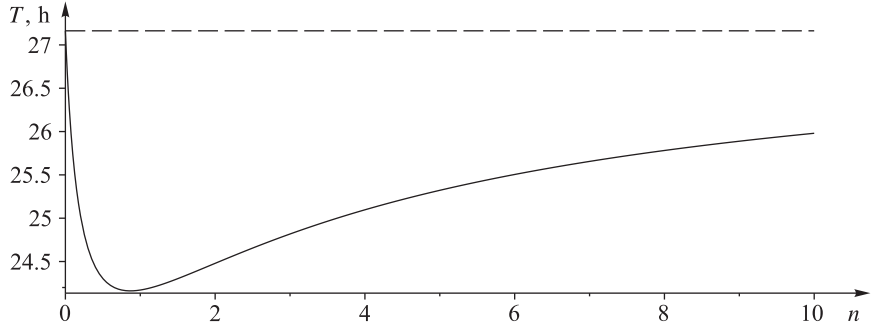}
\caption{The dependence of
period~$T$ on $n$ at the equator for $\langle \rho \rangle
=5.51~\mbox{g}/\mbox{cm}^3$ and $\frac{\rho_0}{\rho_1}=5$} \label{f5}
\end{figure}

Further, we shall define $\rho_0$ and $\alpha$ from various known data for the Earth:

{\it We are given the average density of the body
$\langle \rho\rangle =5.51~\mbox{g}/\mbox{cm}^3$ and the ratio between the densities on
the surface and at the center $\frac{\rho_0}{\rho_1}=5$.} In this case
$\rho_0$ and $\alpha$ are defined by (\ref{biz19_}), and the dependence of
the period of revolution at the equator $T$ on $n$ is shown in Fig.~\ref{f5}.

As can be seen in Fig. \ref{f5}, $T(n)$ reaches the minimum at the point $T(0.8675)=24.1610$~h.

\begin{figure}[!ht]
\centering\includegraphics{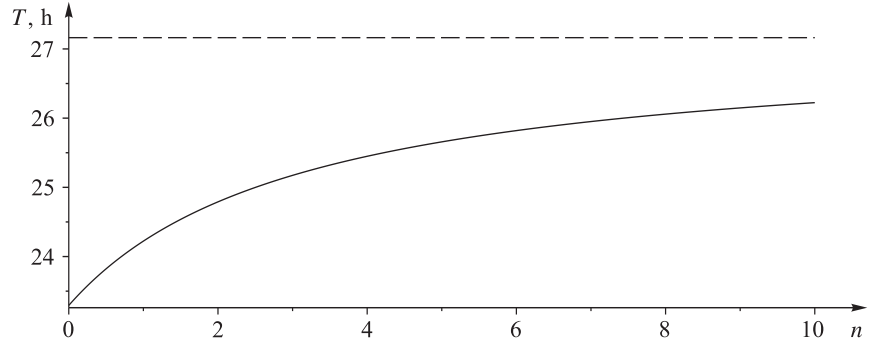}
\caption{The dependence of
period~$T$ on $n$ at the equator for $\langle \rho \rangle
=5.51\mbox{g}/\mbox{cm}^3$ and $\varepsilon=2.5$} \label{f6}
\end{figure}

{\it We are given the average density of the body, $\langle \rho
\rangle =5.51~\mbox{g}/\mbox{cm}^3$, and the ratio between the density on the
surface and the average density,
$\frac{\rho_1}{\langle \rho\rangle }=\varepsilon=2.5$.} The dependence of the period
of revolution at the equator $T$ on $n$ is shown in Fig.~\ref{f6}.

\begin{figure}[!ht]
\centering
\parbox[t]{120mm}{\includegraphics{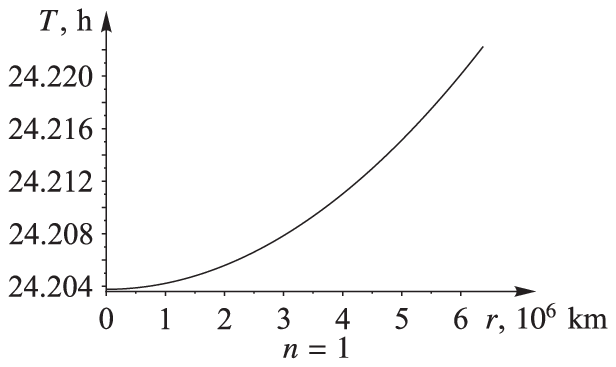}\hfill
\includegraphics{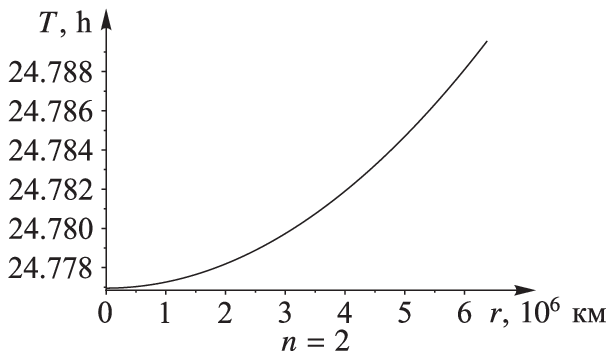}} \\[2mm]
\centering\includegraphics{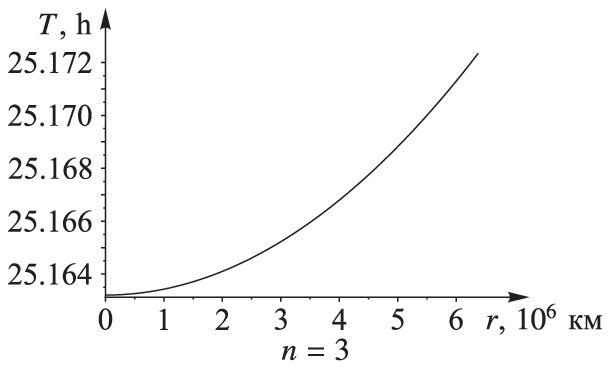}
\caption{The dependence of period~$T$ on the polar radius on the surface
of the inhomogeneous spheroid $\langle \rho \rangle
=5.51\mbox{g}/\mbox{cm}^3$ and $\varepsilon=2.16$ for $n=1$, $n=2$ and
$n=3$.}\label{ff4}
\end{figure}

The dependence of the period of revolution $T$ on the polar radius $r$ on
the surface is shown in Fig. \ref{ff4}.

\section{Figures of equilibrium in $S^3$}

One of the generalizations of the above results is that they are carried
over to the spaces of constant curvature $S^3$ and $L^3$, by analogy with
celestial mechanics of point masses~\cite{we,killing,kh,shred}. There is
a~vast classical and recent literature on the dynamics of gravitating
point masses (see~\cite{Al,we,BM01,BMK01}, in which, for example, the
well-known analogs of the Kepler law and those of the three-body problem
were studied. However, a~particular generalization of the theorems of
Newtonian potential to $S^3$ and $L^3$ was performed only in~\cite{ka}. As
will be shown below, in this case the problem of equilibrium figures
becomes considerably more complex. In particular, even in the case of
homogeneous ellipsoids the rigid body rotation of a~fluid mass is
impossible (we recall that an ellipsoid in curved space is said to be
a~body obtained by the intersection of the sphere $S^3$ or the Lobachevsky
space $L^3$, embedded in ${\mathbb R}^4$, with a~conical quadric). One of
the difficulties is due to the fact that although some generalizations of
Ivory's theorem on the potential of the elliptic layer~\cite{ka} are
possible, this and similar theorems cannot be completely extended to $S^3$
and $L^3$ (they are closely related to the homogeneity of plane space).

\smallskip\begin{small}{\bf Remark.}
Generalizations of the problem of equilibrium figures to the relativistic
case are also possible, see, e.g., the review~\cite{maknp}. Unfortunately,
attempts to obtain explicit analytical exact solutions along these lines
have yielded no results so far. This direction is a~new research area.

\end{small}\smallskip

\subsection{Steady-state axisymmetric solutions in $S^3$}

To explore possible figures of equilibrium in $S^3$, we choose curvilinear
coordinates, as was done for the plane space $E^3$. For convenience, we
assume $S^3$ to be embedded into $E^4$, then the transition to the
coordinates under consideration has the form
$$
x_0=\pm\sqrt{R^2-r^2-Z^2(r,\mu)}, \ x_1=Z(r,\mu), \ x_2=r\cos(\varphi), \ x_3=r\sin(\varphi),
$$
where $Z(r,\mu)$ is defined, as before, by the specific problem statement. The metric tensor can be represented as
$$
{\bf G}=\begin{pmatrix}
g_{11}& g_{12} & 0 \\
g_{12}& g_{22} & 0 \\
0 & 0 & r^2 \\
\end{pmatrix},
$$
where
$$
g_{11}=1-Z_{r}^2+\frac{(r+ZZ_r)^2}{R^2-r^2-Z^2}, \quad
g_{12}=\frac{Z_{\mu}(rZ+(R^2-r^2)Z_r)}{R^2-r^2-Z^2}, \quad
g_{22}=\frac{(R^2-r^2)Z_{\mu}^2}{R^2-r^2-Z^2}.
$$

We shall seek a~steady-state solution for which the velocity distribution
of fluid particles has the form
$$
\dot{r}=0, \quad  \dot{\mu}=0, \quad  \dot{\varphi}=\omega(r,\mu).
$$
As above, assuming that the density depends only on
$\mu$ and using the equations of Section~\ref{subsec1_1}, we obtain the system
\begin{equation}
\label{b1}
\begin{gathered}
\frac{\partial U}{\partial r}+\frac{1}{\rho}\frac{\partial p}{\partial r}=r\omega^2, \quad
    \frac{\partial U}{\partial \mu}+\frac{1}{\rho}\frac{\partial p}{\partial \mu}=0, \\
\triangle_{r\mu}U=4 \pi G\rho(\mu),\\
\triangle_{r\mu}=\frac{x_0}{rZ_{\mu}}\frac{\partial}{\partial r}
    \left[\frac{rZ_{\mu}}{x_0}\left(1-\frac{r^2}{R^2} \right)\frac{\partial}{\partial r}\right]
    +\frac{x_0}{Z_{\mu}}\frac{\partial}{\partial \mu}
    \left[\frac{1}{x_0Z_{\mu}}\left(1+Z_r^2-\frac{(Z-rZ_r)^2}{R^2}\right)\frac{\partial}{\partial \mu}\right]+\\
+\frac{x_0}{rZ_{\mu}}\left(
    \frac{\partial}{\partial r}
    \left[\frac{r}{x_0}\left(Z_r+\frac{r(Z-rZ_r)}{R^2}\right)\frac{\partial}{\partial\mu}\right]+
\frac{\partial}{\partial r}
    \left[\frac{r}{x_0}\left(Z_r+\frac{r(Z-rZ_r)}{R^2}\right)\frac{\partial}{\partial\mu}\right]
\right),
\end{gathered}
\end{equation}
where $x_0=\sqrt{R^2-r^2-Z^2(r,\mu)}$ and it is assumed that the density
$\rho(\mu)$ vanishes everywhere outside the body
($\mu_0<\mu$), and at the free boundary $\mu=\mu_0$ the
pressure is zero as well:
$$
p(r,\mu)|_{\mu=\mu_0}=0.
$$

As we can see, the hydrodynamical equations remain the same
as in $E^3$. Therefore, as in Section~\ref{sec3}, their solution inside
the region ($\mu\le\mu_0$) filled with fluid can be represented as
\begin{equation}
\label{eq37_12}
\begin{gathered}
p(r,\mu)=\rho_0U(r,\mu_0)-\rho(\mu) U(r,\mu) +
    \int\limits_{\mu_0}^{\mu} U(r,\mu)\,\frac{d\rho(\mu)}{d\mu}\,d\mu,\quad \rho_0=\rho(\mu_0),\\
\omega^2(r,\mu)=\frac{1}{r\rho(\mu)}\left(\rho_0\,\frac{d U}{d r}(r,\mu_0)+
    \int\limits_{\mu_0}^{\mu} \frac{d U}{d r}(r,\mu)\,\frac{d\rho(\mu)}{d\mu}\,d\mu\right).
\end{gathered}
\end{equation}

\subsection{A homogeneous spheroid in $S^3$}

We now consider in more detail the case of a~homogeneous spheroid, when
for $\mu\le\mu_0$ the density $\rho(\mu)=\rho_0={\rm const}$. The
generalization of confocal stratification in $S^3$ is given as follows
(cf. Section \ref{subsec2.2}):
$$
\frac{x_0^2}{R^2-d^2\mu^2}-\frac{x_1^2}{d^2\mu^2}-\frac{x_2^2+x_3^2}{d^2(1+\mu^2)}=0,\quad
\mu\in\bigg[0,\frac{R}{d}\bigg].
$$
Hence, we obtain
$$
Z(r,\mu)=\pm\sqrt{d^2\mu^2 - r^2\frac{R^2+d^2}{R^2} \frac{\mu^2}{1+\mu^2}}.
$$

As in the previous case (see Section~\ref{subsec2.2}), the parameter $d$
and the boundary $\mu_0$ of a~liquid spheroid with semi-axes $a$ and $b$
are given by
$$
d=\sqrt{a^2-b^2},\quad \mu_0=\frac{b}{\sqrt{a^2-b^2}}.
$$

According to~(\ref{eq37_12}), in the case of a~homogeneous spheroid
$\frac{d\rho}{d\mu}=0$, therefore, the angular velocity of the fluid depends only on $r$:
\begin{equation}
\label{eq38_12}
\omega^2(r)=\frac{1}{r}\,\frac{\partial U}{\partial r}(r, \mu_0).
\end{equation}

We shall seek solutions to the equation for the potential~(\ref{b1}) in the form of a~power series
in the parameter $\frac{d^2}{R^2}$:
$$
U(r,\mu)=2\pi Gd^2\sum\limits_{n=0}^{\infty}\left(\frac{d}{R}\right)^{2n} U_n(r,\mu).
$$
As can be shown, all terms of this series are polynomials in $r$. It is convenient to represent them as
$$
U_n(r,\mu)=\sum\limits_{n=0}^{\infty}\left(\frac{r}{d}\right)^{2m} \frac{u_{n,\mu}(\mu)}{2^m(1+\mu^2)^m}.
$$
The potential $U_0(r,\mu)$ is equal (up to a~multiplier) to the potential
of the Maclaurin spheroid (see Section~\ref{sec2.3}):
$$
U_0(r,\mu)=u_{0,0}(\mu)+\frac{r^2}{d^2}\,\frac{u_{0,1}(\mu)}{2(1+\mu^2)},
$$
inside the spheroid ($\mu\le \mu_0$):
$$
\begin{gathered}
u^{\rm in}_{0,0}(\mu)=\rho_0(1+\mu_0^2)\big(\mu^2 - \mu_0(1+\mu^2)\arcctg \mu_0\big),
\\
u^{\rm in}_{0,1}(\mu)=\rho_0\Big( \mu_0(1+3\mu^2)\big((1+\mu_0^2)\arcctg \mu_0 - \mu_0\big) - 2\mu^2\Big),
\end{gathered}
$$
outside the spheroid ($\mu_0<\mu$):
$$
\begin{gathered}
u^{\rm out}_{0,0}(\mu)=\rho_0\mu_0(1+\mu_0^2)\big(\mu - (1+\mu^2)\arcctg \mu\big),
\\
u^{\rm out}_{0,1}(\mu)=\rho_0\mu_0(1+\mu_0^2)\big( (1 + 3\mu^2)\arcctg \mu - 3\mu\big).
\end{gathered}
$$

\begin{figure}[!ht]
\centering\includegraphics{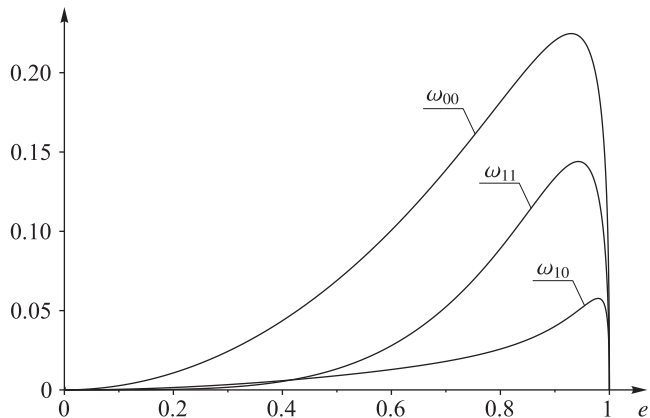}
\caption{Dependences of $\omega_{00}$, $\omega_{11}$, and $\omega_{10}$ on the eccentricity $e$.}
\label{fb1}
\end{figure}

We shall assume that the space curvature is very small ($R^2\gg a^2$)
and, therefore, restrict ourselves to calculating the first correction
$$
U_1(r,\mu) =\frac{r^4}{d^4}\frac{u_{1,2}(\mu)}{4(1+\mu^2)^2}
    +\frac{r^2}{d^2}\frac{u_{1,1}(\mu)}{2(1+\mu^2)}+u_{1,0}(\mu),
$$
where the functions $u_{1,0}(\mu)$, $u_{1,1}(\mu)$,
and $u_{1,2}(\mu)$ satisfy the equations
\begin{equation}\small
\label{eq10_12-}
\begin{gathered}
\frac{d}{d\mu}\bigg((1+\mu^2)\frac{d u_{1,2}}{d\mu}\bigg) - 20u_{1,2} + 16u_{0,1}=0, \\
\frac{d}{d\mu}\bigg((1+\mu^2)\frac{d u_{1,1}}{d\mu}\bigg) - 6u_{1,1} - \mu(1+\mu^2)\frac{du_{0,1}}{d\mu}
- 6(2+\mu^2)u_{0,1}+8 u_{1,2} + 4 \rho_0(1+\mu^2)=0, \\
\frac{d}{d\mu}\bigg((1+\mu^2)\frac{d u_{1,0}}{d\mu}\bigg) - 2u_{1,1} - \mu(1+\mu^2)
\frac{du_{0,0}}{d\mu} + 2\mu^2( u_{0,1} + \rho_0(1+\mu^2))=0.
\end{gathered}
\end{equation}
The functions $u_{1,0}$, $u_{1,1}$, and $u_{1,2}$ must also satisfy the following boundary conditions:
$$
\begin{gathered}
\frac{d u_{1,m}^{\rm in}}{d \mu}\bigg|_{\mu=0}=0, \quad  m=0,1,2.
\\
u_{1,m}^{\rm in}|_{\mu=\mu_0}=u_{1,m}^{\rm out}|_{\mu=\mu_0}, \quad
\frac{d u_{1,m}^{\rm in}}{d\mu}\bigg|_{\mu=\mu_0}=\frac{d u_{1,m}^{\rm out}}{d\mu}\bigg|_{\mu=\mu_0}, \quad  m=0,1,2.
\\
U_1(r,\mu)\big|_{\mu=\frac{R}{d}}= O(R^2).
\end{gathered}
$$
Since the solution of the resulting system is rather unwieldy, we omit it
here and confine ourselves to the expression for the angular velocity of
the fluid, for which, according to~(\ref{eq38_12}), we find
$$
\frac{\omega^2(r)}{2 \pi G}=\frac{u_{0,1}^{\rm in}(\mu_0)}{1 + \mu_0^2} +
    \frac{1}{R^2}\left( \frac{u_{1,2}^{\rm in}(\mu_0)}{(1 + \mu_0^2)^2}r^2+
    \frac{u_{1,1}^{\rm in}(\mu_0)}{1 + \mu_0^2}d^2  \right)  + O\left(\frac{d^4}{R^{4}}\right).
$$
Substituting the solution for $u_{1,m}^{\rm in}(\mu_0)$ and expressing $\mu_0$
in terms of the eccentricity of the boundary using the formula $e=\frac{1}{\sqrt{1+\mu_0^2}}$,
we obtain an explicit representation for the angular velocity in the form
$$
\begin{gathered}
\frac{\omega^2(r)}{2\pi G\rho_0}=\omega_{00} +
\frac{1}{R^2}\left(\omega_{11}r^2 + \omega_{10}a^2\right) +
O\left(\frac{d^2}{R^{4}}\right),
\\
\omega_{00}=-\frac{\sqrt{1 - e^2}}{e}\left(2 -
\frac{3}{e^2}\right)\arcsin e - \frac{3}{e^2}(1 - e^2),
\\
\omega_{11}=-\frac{\sqrt{1 - e^2}}{e}\left(12 - \frac{30}{e^2} +
\frac{35}{2e^4}\right)\arcsin e + \left(\frac{4}{3} -
\frac{55}{3e^2} + \frac{35}{2e^4}\right)(1 - e^2),
\\
\omega_{10}=\frac{\sqrt{1 - e^2}}{e}\left(16 - \frac{27}{2e^2}
+\frac{10}{e^4}\right)\arcsin e - \left(\frac{1}{3} -
\frac{41}{6e^2} + \frac{10}{e^4}\right)(1 - e^2),
\end{gathered}
$$
where we have also passed from the parameter $d$ (which tends to zero as
$e\to0$) to the value of the largest principal semi-axis $a$.
The graphs of dependence of each of the corrections for the angular velocity on the eccentricity
is presented in Fig.~\ref{fb1}.

\smallskip
\textit{Thus, in the space of constant (positive) curvature the
homogeneous liquid self-gravitating spheroid cannot rotate as  a~rigid
body, and the angular velocity distribution of fluid particles depends
only on the distance to the symmetry axis: $\omega=\omega(r)$.}

\smallskip\begin{small}{\bf Remark.}
For completeness we also present the equations which describe axisymmetric
figures of equilibrium in curvilinear orthogonal coordinates
$(\mu,\nu,\varphi)$ and are defined
 as follows:
$$
\begin{gathered}
\frac{x_0^2}{d^2}=\frac{(\delta -\mu)(\delta +\nu)}{\delta +1},\quad \frac{x_1^2}{d^2}=\mu\nu,\quad
    \delta =\frac{R^2}{d^2}    \\
\frac{x_2^2}{d^2}=\frac{(1+\mu)(1-\nu)}{\delta +1}\cos^2\varphi,\quad
    \frac{x_3^2}{d^2}=\frac{(1+\mu)(1-\nu)}{\delta +1}\sin^2\varphi,\quad
    0<\mu<\delta ,\quad 0<\nu<1.
\end{gathered}
$$
In this case the system~(\ref{b1}) takes the form
$$
\begin{gathered}
\frac{\partial U}{\partial \mu}+\frac{1}{\rho(\mu)}\,\frac{\partial p}{\partial \mu}=
    -\frac{\delta d^2}{2(\delta +1)}(1-\nu)\omega^2,\quad
\frac{\partial U}{\partial \nu}+\frac{1}{\rho(\mu)}\,\frac{\partial p}{\partial \nu}=
    \frac{\delta d^2}{2(\delta +1)}(1+\mu)\omega^2,\\
\Delta_{\mu\nu}U(\mu,\nu)=4\pi G\rho(\mu),\\
\begin{aligned}
R^2 \Delta_{\mu\nu}&=\frac{4}{\mu+\nu}
    \bigg( \sqrt{\mu(\delta -\mu)}\,\frac{\partial}{\partial \mu}\bigg((1+\mu)\sqrt{\mu(\delta -\mu)}\,\frac{\partial}{\partial \mu}\bigg)
    +\\
&+\sqrt{\nu(\delta +\nu)}\,\frac{\partial}{\partial \nu}\bigg((1-\nu)\sqrt{\nu(\delta +\nu)}\,\frac{\partial}{\partial \nu}\bigg) \bigg).
\end{aligned}
\end{gathered}
$$

This form of equations is preferable if it is necessary to
obtain a~solution in terms of quadratures (and not in the form of a~power
series).

\end{small}\smallskip

\section{Discussion}

Thus, in this paper we have systematically analyzed the problem
of\linebreak inhomogeneous axisymmetric equilibrium figures of an ideal
self-gravitating fluid. We have obtained the most general solution
describing a~stratified spheroid (the angular velocity of the fluid takes
the same value on the layer with equal density, i.e.,
$\omega=\omega(\mu)$). This solution naturally yields the above-mentioned
spheroids with piecewise constant density
distribution~\cite{esteban,mms}. It is shown that the angular
velocity of the outer surface of the spheroid with confocal stratification
of density $\rho$ is the same as that of the homogeneous Maclaurin
spheroid with density $\langle \rho\rangle $. Therefore, this model cannot
be used to explain the deviation of the compression of planets from the
compression of the Maclaurin spheroids rotating with the same  angular
velocity.

We have also presented a~fairly detailed review (and a~formulation in
modern terms) of results in this vein. Of special note is Chaplygin's work
(previously unpublished and found in archives) on spheroids with homofocal
density\linebreak stratification.

In the last section we have considered the problem of the conditions for
equilibrium of a~homogeneous spheroid in the spaces of constant curvature
$S^3$ and shown that in this case the fluid cannot rotate as a~rigid body
and that the angular velocity of fluid particles depends only on the
distance to the symmetry axis: $\omega=\omega(r)$.

We conclude by pointing out some open problems
related to possible\linebreak generalizations of the above results.

\textbf{1.} The stratified analogs of the Maclaurin spheroids raise the
question of their stability. Of particular importance is here in all
probability their secular stability, which was considered by
Lyapunov~\cite{lyapunov} for the case of homogeneous fluid density. In his
analysis of the perturbation of the free surface by spherical harmonics he
concluded that the higher the order of a~harmonic, the larger the value of
eccentricity at which the loss of secular stability occurs. In the general
case Lyapunov arrived at the conclusion that the secular stability of the
Maclaurin spheroids is lost under arbitrary deformations if the
eccentricity becomes equal to 0.8126 (for the special case of ellipsoidal
perturbations this result was obtained by Dirichlet~\cite{ell36}).

As far as this problem is concerned, no finite-dimensional equations
governing the dynamics of stratified ellipsoids have been obtained so far
(see~\cite{ell36,Fasso,Rim}). Because of this it is difficult to obtain
all sufficient stability criteria determined by the finite dimensionality
of the system (Lyapunov's theorem, KAM theory).

\textbf{2.} Historically, attempts to derive the first equations of
stratified ellipsoids go back to~\cite{betti}, but, as
Tedone~\cite{tedone} noted, Betti made a~mistake in his study. In this
connection, the question of possible existence of three-axial
inhomogeneous ellipsoids still awaits its solution.

\textbf{3.} The above solution for spheroids with confocal stratification
is evidently the only solution possible, for which $\omega=\omega(\mu)$,
but no proof of this fact has been found.

\textbf{4.} The problem of stability of the found figures of equilibrium
with respect to both ellipsoidal and arbitrary perturbations is also an
open question.

\textbf{5.} Another interesting problem is that of obtaining an explicit
solution (not in the form of a~power series) for a~homogeneous spheroid in
curved space and the search for other possible figures of equilibrium in
the spaces of constant curvature.

\textbf{6.} We recall that for the Maclaurin and Jacobi ellipsoids there
exists\linebreak a~``dynamical'' generalization, due to Dirichlet, where the
self-gravitating liquid ellipsoid retains an ellipsoidal shape but changes
the directions and dimensions of the semi-axes during its motion.
It is unknown whether there exists such a~dynamical
generalization for inhomogeneous figures of equilibrium.

\smallskip\begin{small}{\bf Remark.}
A simple extension of Dirichlet's method, for example, to a~family with
confocal density stratification (see Section~\ref{sec2}) is impossible
since in Dirichlet's solution the same fluid particles move in ellipsoids
forming at each instant of time a~homothetic (and not confocal) foliation.

\end{small}\smallskip

\textbf{7.} Another possible generalization involves finding the figures
of equilibrium of a~stratified gas cloud. In this case, in order to close
the system~(\ref{eq_zv}), one uses, as a~rule, thermodynamical equations
(for applications to fluid mass dynamics see the review~\cite{bmk} and
references therein). In particular, one of the simplest assumptions used
in~\cite{dyson} is that the temperature of the fluid/gas is constant along
the entire volume $T(r,\mu)=T_0={\rm const}$. In the case of an ideal gas this
leads to a~linear relation between density and pressure
\begin{equation}
\label{eq_5_1}
p=\lambda\rho,\quad \lambda=RT_0,
\end{equation}
where $R$~is the universal gas constant.

Assuming that $\rho=\rho(\mu)$, we obtain from~(\ref{eq_zv})
and~(\ref{eq_5_1})
the system
$$
\begin{gathered}
\frac{\partial U}{\partial r}=r\omega^2,\quad  \frac{\partial U}{\partial \mu}=\frac{\lambda\rho'(\mu)}{\rho(\mu)},\\
\Delta_{r\mu}^{}U=4\pi G\rho(\mu).
\end{gathered}
$$
One of the unknowns in these equations is the function $Z(r,\mu)$
characterizing possible equilibrium figures of the cloud of an ideal gas.

\smallskip\begin{small}{\bf Remark.}
To close the system, one can use, instead of the equation of
state~(\ref{eq_5_1}), the condition that the fluid flow be barotropic.

\end{small}\smallskip

\begin{small}
\textbf{Acknowledgements.} The authors thank A.\,Albouy for useful advice and invaluable assistance in the course of work.
\end{small}


\begin{thebibliography}{99}
\small\itemsep=-2pt

\bibitem{Al} Albouy, A.: There is a~Projective Dynamics. Eur. Math. Soc.
   Newsl. (89), 37--43 (2013)

\bibitem{appel} Appell, P.: Trait\'e de M\'ecanique Rationnelle:
T.\,4-1.~Figures d'\'Equilibre d'une Masse liquide Homog\`ene en Rotation. Gautier-Villars, Paris (1921)

\bibitem{betti}
Betti E.
Sopra i moti he onservano la gura ellissoidale a~una massa uida
    eterogenea.
Annali di Matemati a~Pura ed Appli ata,  Serie~II,
{\bf X}, 173187 (1881)

\bibitem{PS}  Borisov, A.\,V., Mamaev, I.\,S.: Poisson Structures and Lie
   Algebras in Hamiltonian Mechanics. Izd. UdSU, Izhevsk (1999)
   (in Russian)

\bibitem{we} Borisov, A.\,V., Mamaev, I.\,S.: The Restricted Two-Body
   Problem in Constant Curvature Spaces. Celestial Mech. Dynam. Astronom.
   {\bf 96}(1), 1--17 (2006)

\bibitem{BM01} Borisov, A.\,V., Mamaev, I.\,S.: Relations Between
   Integrable Systems in Plane and Curved Spaces. Celestial Mech. Dynam. Astronom. {\bf 99}(4), 253--260 (2007)

\bibitem{BMK01} Borisov, A.\,V., Mamaev, I.\,S., Kilin, A.\,A.: Two-Body
   Problem on a~Sphere. Reduction, Stochasticity, Periodic Orbits. Regul.
   Chaotic Dyn. {\bf 9}(3), 265--279 (2004)

\bibitem{bmk} Borisov, A.\,V.,  Mamaev, I.\,S.,  Kilin, A.\,A.:  The Hamiltonian
   Dynamics of Self-gravitating Liquid and Gas Ellipsoids. Regul.
   Chaotic Dyn. {\bf 14}(2), 179--217 (2009)

\bibitem{ell30} Chandrasekhar,~S.:  Ellipsoidal Figures of Equilibrium.
   Yale University Press, New Haven (1969)

\bibitem{chaplygin} Chaplygin, S.\,A.: Steady-State Rotation of a~Liquid
    homogeneous spheroid  In Collected works: Vol.\,2.~Hydrodynamics.
    Aerodynamics. Gostekhizdat, Moscow (1948)

\bibitem{clero} Clairaut, A.\,C.: Th\'{e}orie de la Figure de la Terre: Tir\'ee
   des Principes de l'Hydrostratique. Paris Courcier, Paris (1743)

\bibitem{craik} Craik, A.\,D.\,D.: James Ivory's Last Papers on the `Figure
   of the Earth' (with biographical additions). Notes Rec. R. Soc.
   Lond. {\bf 56}(2), 187--204 (2002)

\bibitem{ell36} Dirichlet, G.\,L.: Untersuchungen \"uber ein Problem der
   Hydrodynamik (Aus dessen Nachlass hergestellt von Herrn R.\,Dedekind zu
   Z\"urich). J. Reine Angew. Math. (Crelle's Journal) {\bf 58}, 181--216 (1861)

\bibitem{dyson} Dyson, F.\,J.: Dynamics of a~Spinning Gas Cloud. J. Math.
   Mech. {\bf 18}(1), 91--101 (1968)

\bibitem{esteban} Esteban, E.\,P., Vasquez, S.: Rotating Stratified
    Heterogeneous Oblate Spheroid in Newtonian Physics. Celestial Mech.
    Dynam. Astronom. {\bf 81}(4), 299--312 (2001)

\bibitem{Fasso} Fass\`{o}, F., Lewis, D.: Stability properties of the
    Riemann ellipsoids. Arch. Ration. Mech. Anal. {\bf 158}, 259--292
    (2001)

\bibitem{ferrers} Ferrers, N.\,M.: On the Potentials, Ellipsoids,
    Ellipsoidal Shells, Elliptic Laminae, and Elliptic Rings, of Variable
    Densities. Quart. J. Pure Appl. Math. {\bf 14}, 1--23 (1875)

\bibitem{gaffet} Gaffet, B.: Spinning Gas Clouds: Liouville
Integrability. J. Phys. A: Math. Gen. {\bf 34},  2097--2109 (2001)

\bibitem{hamy} Hamy, M.: \'{E}tude sur la Figure des Corps C\'{e}lestes.
    Ann. de l'Observatoire de Paris. Memories {\bf 19}, 1--54 (1889)

\bibitem{carl_jacobi} Jacobi, C.\,G.\,J.: \"Uber die Figur des
    Gleichgewichts, Poggendorff Annalen der Physik und Chemie, {\bf 33},
    229--238 (1834)

\bibitem{killing} Killing, H.\,W.: Die Mechanik in den Nichteuklidischen
   Raumformen. J.~Reine Angew. Math. {\bf XCVIII}(1), 1--48 (1885)

\bibitem{kkp} Kochin, N.\,E., Kibel, I.\,A., Rose, N.\,V. Theoretical
    Hydromechanics (in Russian), Vol.~1. Fizmatgiz, Moscow (1963)

\bibitem{kzs} Kong, D., Zhang, K., Schubert, G.: Shapes of Two-Layer Models of
   Rotating Planets. J. Geophys. Res. {\bf 115}(E12), doi:10.1029/2010JE003720 (2010)

\bibitem{ka}  Kozlov, V.\,V.: The Newton and Ivory Theorems of Attraction
    in Spaces of Constant Curvature. (Russian)
    Vestnik Moskov. Univ. Ser. I Mat. Mekh. (5), 43--47 (2000)

\bibitem{kh} Kozlov, V.\,V., Harin, A.\,O.: Kepler's Problem in Constant
   Curvature Spaces. Celestial Mech. Dynam. Astronom. {\bf 54}(4) 393--399 (1992)

\bibitem{lihten} Lichtenstein, L.: Gleichgewichtsfiguren Rotierender
   Fl\"ussigkeiten. Springer, Berlin (1933)

\bibitem{liou}
Liouville, J.: Sur la Figure d'une Masse Fluide Homog\`ene, en
E'quilibre et Dou\'ee d'un Mouvement de Rotation. J. de
l'\'Ecole Polytechnique {\bf 14}, 289--296 (1834)

\bibitem{lyapunov}
Lyapunov, A.\,M.: Collected Works, vol.~3. Moscow (1959)

\bibitem{lyttleton} Lyttleton, R.\,A.: The Stability of Rotating Liquid
   Masses. Cambridge University Press, Cambridge (1953)

\bibitem{maclar} MacLaurin, C.: A Treatise of Fluxions: In Two Books.
   Printed by T.W. \& T. Ruddimans, Edinburgh (1742)

\bibitem{maknp} Meinel, R., Ansorg, M., Kleinwachter, A., Neugebauer, G.,
   Petroff, D.: Relativistic Figures of Equilibrium. Cambridge
   University Press, Cambridge (2008)

\bibitem{mms} Montalvo, D., Mart\'{\i}nez, F.\,J., Cisneros, J.: On
   Equilibrium Figures of Ideal Fluids in the Form of Confocal Spheroids
   Rotating with Common and Different Angular Velocities.  (1982)

\bibitem{pizzetti} Pizzetti, P.: Principii della Teorii Meccanica della
   Figura dei Pianeti. Enrico Spoerri, Libraio-Editore, Pisa (1913)

\bibitem{rhdb} Rambaux, N., Van Hoolst, T., Dehant, V., Bois, E.:
    Inertial Core-Mantle Coupling and Libration of Mercury. Astronom.
    Astrophys. {\bf 468}, 711--179 (2007)

\bibitem{Rim} Riemann, B.: Ein Beitrag zu den Untersuchungen \"uber die
   Bewegung eines Fl\"ussigen gleichartigen Ellipso\"ides. Abh. d.
   K\"onigl. Gesel l. der Wiss. zu G\"ottingen (1861)

\bibitem{shred} Schr\"odinger, E.: A Method of Determining
   Quantum-Mechanical Eigenvalues and Eigenfunctions. Proc. Roy. Irish
   Acad. Sect. A {\bf 46}, 9--16 (1940)

\bibitem{tedone}
Tedone O.
Il moto di un ellissoide fluido se ondo
    l'ipotesi di Dirihlet. Annali del la S uola Normale Superiore di Pisa,
{\bf 7}, I--IV+1--100 (1895)

\bibitem{veronnet}
V\'{e}ronnet A.:
Rotation de l'Ellipsoide H\'{e}t\'{e}rog\`{e}ne et Figure Exacte de la
Terre. J.~Math. Pures et Appl., S\'{e}r.\,6 {\bf 8}, 331--463 (1912)

\bibitem{volterra} Volterra, V.: Sur la Stratification d'une Masse Fluide en
   Equilibre. Acta Math. {\bf 27}(1), 105--124 (1903)

\bibitem{david} Williams, D.\,R.: Earth Fact Sheet.
   Structural Geology of the Earth's Interior: Proc. Natl.
   Acad. Sci. {\bf 76}(9), (NASA (17 Nov 2010))

\end{thebibliography}
\end{document}